\documentclass[floats,aps,showpacs,amsmath,prl,twocolumn]{revtex4-1}
\usepackage{graphicx}
\usepackage{color}
\usepackage{amsmath}
\usepackage{rotating}
\usepackage{dcolumn}
\usepackage{bm}

\begin{document}
\newcommand{\g}{\bf}

\title{Quasilinear spin voltage profiles in spin thermoelectrics}

\author{Tamara S.\ Nunner}
\affiliation{\mbox{Dahlem Center for Complex Quantum Systems and Fachbereich Physik, Freie Universit\"at
  Berlin, 14195 Berlin, Germany}}
\author{Felix von Oppen}
\affiliation{\mbox{Dahlem Center for Complex Quantum Systems and Fachbereich Physik, Freie Universit\"at
  Berlin, 14195 Berlin, Germany}}

\date{\today}

\begin{abstract}
Recent experiments show that spin thermoelectrics is a promising approach to generate spin voltages. While spin chemical potentials are often limited to a surface layer of the order of the spin diffusion length, we show that thermoelectrically induced spin chemical potentials can extend much further in itinerant ferromagnets with paramagnetic impurities. In some cases, conservation laws, e.g., for a combination of spin and heat currents, give rise to a linear spin voltage profile. More generally, we find quasilinear profiles involving a spin thermoelectric length scale which far exceeds the spin diffusion length.
\end{abstract}

 \maketitle

{\it Introduction.}---The field of spintronics~\cite{Spintronics,Fert} is fueled by the vision of more efficient microelectronic devices which exploit the spin degree of freedom $\sigma$ of electrons in addition to their charge $e$. To date, attention focused largely on electric phenomena. Recently, much consideration is also given to spin-dependent thermoelectric effects which opened up the new research field of spin caloritronics~\cite{SpinCaloritronics}. A potential milestone of this emerging field were the recent observations of the spin Seebeck effect in ferromagnets~\cite{Uchida,Jaworski,Uchida2}, where a spin voltage develops in response to an applied temperature gradient. This could be valuable for spintronics applications as the operating principle of a spin battery. Indeed thermally driven spin injection has recently been realized experimentally~\cite{Slachter}.

Besides such practical implications, the spin Seebeck effect also raises theoretical puzzles. Remarkably, the spin chemical potential is observed to vary (quasi)linearly along the sample~\cite{Uchida,Jaworski,Uchida2}. This is surprising since unlike charge, the electronic spin is generally not conserved due to spin-orbit interactions or magnetic impurities. While charge conservation implies that the chemical potential drops linearly, spin non-conservation would seem to imply that the spin chemical potential should vary exponentially on the scale of the spin diffusion length~\cite{Hatami}. 

Proposed mechanisms for the linear spin-voltage profile are based on spin pumping between ferromagnet and contacts~\cite{Xiao,Adachi, Adachi2}, with
previous theoretical works~\cite{Xiao,Adachi, Adachi2,Tulapurkar} focusing on ferromagnetic insulators or magnon drag. In this paper, we consider the contribution of the conduction electrons to the spin Seebeck effect in a ferromagnetic metal. In particular, we discuss the question under what circumstances a (quasi)linear variation of the spin chemical potential can be obtained. Although it is well known that elastic spin-flip scattering leads to an exponential decay of the spin chemical potential near the edges of the sample~\cite{Hatami}, we find that the situation changes qualitatively when additional inelastic spin-flip scattering is present as, e.g., arising from the exchange interaction with magnetic impurities.

As a model system, we consider inelastic spin-flip scattering effected by the exchange interaction with partially spin-polarized (paramagnetic) impurities. In the absence of elastic spin relaxation of the conduction electrons or without additional relaxation mechanisms for the impurity spins, a truly linear dependence of the spin voltage can be recovered due to the interplay of a conservation law for a combination of spin and heat currents with the spin asymmetry inherent to inelastic spin-flip scattering. Remarkably, even in the more general case with additional direct spin relaxation for both conduction electrons and impurities, we find a quasilinear variation of the spin voltage whose characteristic scale is much larger than the spin diffusion length.

Another question raised by the experiment is the origin of the observed spin Seebeck effect \cite{Uchida}. In the absence of response functions which are off-diagonal in spin, linear response relates the spin-resolved charge and heat currents, ${\bf J}_\sigma$ and ${\bf J}_{Q\sigma} $, to the gradients in the spin-resolved chemical potentials and temperatures, $\mu_\sigma$ and $T_\sigma$,  
\begin{eqnarray}
{\bf J}_\sigma &=& -\frac{\sigma_\sigma}{e^2} {\bf \nabla} \mu_\sigma
-\frac{\sigma_\sigma S_\sigma}{e}  {\bf \nabla} T_\sigma \nonumber\\
{\bf  J}_{Q \sigma} &=& -\frac{1}{e} T_\sigma \sigma_\sigma S_\sigma
{\bf \nabla} \mu_\sigma - (\kappa_\sigma+T_\sigma \sigma_\sigma
S_\sigma^2) {\bf \nabla} T_\sigma \,.
\label{eq:LinearResponse}
\end{eqnarray}
Here, $\sigma_\sigma$ denotes the (spin-resolved) conductivity, $S_\sigma$ the thermopower, and $\kappa_\sigma$ the heat conductivity. It is natural to ask whether the observed spin Seebeck effect indeed originates from the spin thermopower $S_-=S_\uparrow - S_\downarrow$. We find that this occurs in special limits but is generically not the case.

{\it Model.}---We consider an itinerant ferromagnet where the exchange interaction polarizes the electron spins through the mean field $h$ and the electron dispersion is spin split, $\epsilon_{\bf p\sigma}=\epsilon_{\bf p}-\sigma h$. Besides elastic potential scattering we assume that the electrons are subject to inelastic spin-flip scattering (with rate $\Gamma_\sigma^{\rm sf}$) on partially polarized magnetic impurities. The electron spin can also relax directly ($\Gamma_\sigma^{\rm dir}$), e.g., via spin-orbit interaction. Similarly, the impurities can relax their spin directly ($\Gamma_i^{\rm dir}$), e.g., via the combined action of spin-orbit interaction and phonons~\cite{Fert95} in addition to spin-flip scattering with the conduction electrons ($\Gamma_i^{\rm sf})$.

Our discussion of the spin thermoelectric properties of the ferromagnet is based on the kinetic equation 
\begin{equation}
\label{eq:Boltzmann}
\{\partial_t +  {\bf v}_{\bf p\sigma}\cdot {\bf \nabla}_{\bf r}  +  {\bf F}\cdot {\bf \nabla}_{\bf p}\}
 n_{\bf p\sigma}= S_{\bf p\sigma}^{\rm el}+S_{\bf p\sigma}^{\rm inel},
\end{equation}
for the electronic distribution function $n_{\bf p\sigma}$. Here, ${\bf v}_{\bf p\sigma}={\bf\nabla}_{\bf p} \epsilon_{\bf p\sigma}$, and ${\bf F}$ is an externally applied force. Elastic scattering and inelastic spin-flip scattering are described through the collision integrals
\begin{eqnarray}
&&S_{\bf p\sigma}^{\rm el}= \sum_{\bf p'\sigma'} 
\!W^{\sigma\sigma'}_{\bf pp'} \delta(\epsilon_{\bf p\sigma}-\epsilon_{\bf p'\sigma'})
\left(n_{\bf p'\sigma'}\!\!-\! n_{\bf p\sigma}\right)
\nonumber \\
&&S_{\bf p\sigma}^{\rm inel}=  \sum_{\bf p'} W^{\rm sf}_{\bf pp'} 
\delta(\epsilon_{\bf p\sigma}+\sigma \epsilon_i-\epsilon_{\bf p'\bar\sigma})
\nonumber \\
&& \quad\quad
\times\left[(1-n_{\bf p\sigma})n_{\bf p'\bar \sigma}g_{\sigma +} - n_{\bf p\sigma}(1-n_{\bf p'\bar \sigma}) g_{\sigma -} \right],
\label{eq:CollisionIntegrals}
\end{eqnarray}
where $g_{\sigma \pm} = g + (1\pm \sigma)/2$. Here, $W^{\sigma\sigma}$ results from elastic impurity scattering and can be spin dependent as, e.g., for spin polarized impurities, $W^{\sigma\bar \sigma}$ denotes the elastic direct spin relaxation, and $W^{\rm sf}$ denotes the strength of the inelastic spin-flip scattering with paramagnetic impurities. For definiteness, we consider isotropic scattering due to pointlike scatterers. Since paramagnetic impurities in a ferromagnet  will be at least partially spin polarized, the impurity $S_z$-levels split by an energy of $\epsilon_i$. We treat the impurity spin as classical, i.e., we describe deviations from the maximal spin orientation with a bosonic distribution function $g$, which is governed by
\begin{equation}
\label{eq:ImpEq}
\frac{\partial g}{\partial t} = \sum_{\bf p}
S_{\bf p\uparrow}^{\rm inel} 
+ \left[ W_{i+}^{\rm dir} (g+1)-W_{i-}^{\rm dir} g \right].
\end{equation}
$W_{i\pm}^{\rm dir}$ denotes direct relaxation of the impurity spin. 

{\it Transport coefficients.}---The transport coefficients $\sigma_\sigma$, $S_\sigma$, and $\kappa_\sigma$ in Eq.~(\ref{eq:LinearResponse}) can be determined as usual by evaluating the spin-resolved charge and heat currents from the Boltzmann equation~(\ref{eq:Boltzmann}). Multiplying Eq.~(\ref{eq:Boltzmann}) by velocity ${\bf v}_{\bf p\sigma}$ and ${\bf v}_{\bf p\sigma} (\epsilon_{\bf p\sigma}-\mu_0)$ (where $\mu_0$ is the equilibrium chemical potential) and summing over momenta, we find $\sigma_\sigma = e^2 N_\sigma D_\sigma$ for the conductivity. Here, $N_\sigma$ is the electronic density of states and $D_\sigma = v_\sigma^2/d \Gamma_\sigma^a$ is the diffusion coefficient in terms of $v_\sigma=v_{{\bf p}_F,\sigma}$ 
and $\Gamma_\sigma^a = \Gamma_\sigma^{\rm tr} + \Gamma_\sigma^{\rm sf}+\Gamma_\sigma^{\rm dir}$, where $\Gamma_\sigma^{\rm tr}$ is the transport rate for spin-conserving elastic scattering, $\Gamma_\sigma^{\rm dir} = N_{\bar \sigma}(\mu_0) W^{\sigma \bar \sigma}$, and $\Gamma_\sigma^{\rm sf} =N_{\bar \sigma} (\mu_0 + \sigma \epsilon_i) ({W^{\rm sf}}/{\sinh \frac{\epsilon_i}{k_B T}})$ the rates for elastic and inelastic spin-flip scattering, respectively. The thermopower and the thermal conductivity obey the Mott relation $S_\sigma = ({\pi^2 k_B^2 T}/{3e}) ({\partial \ln \sigma_\sigma(\mu_0)}/{\partial \mu_0})$ and the Wiedemann-Franz law $\kappa_\sigma= ({\pi^2 k_B^2 T}/{3e^2}) \sigma_\sigma$.

{\it Continuity equations.}---Information about the spatial dependence of (spin) chemical potential and (spin) temperature is contained in the continuity equations for the spin-resolved charge and heat currents. These can be found in the usual manner by a momentum sum over the Boltzmann equation (\ref{eq:Boltzmann}) as well as over Eq.\ (\ref{eq:Boltzmann}) multiplied by $(\epsilon_{\bf p\sigma}-\mu_0)$. We evaluate the resulting expressions by linearizing the collision integrals using $n_{\bf p \sigma} \simeq f_{\bf p \sigma}^0 + \delta n_{\bf p \sigma}$ and $g \simeq g^0(\epsilon_i) + \delta g$, and expanding the scattering rates about $\mu_0$, $\Gamma_{\bf p\sigma} \simeq \Gamma_\sigma+\Gamma_\sigma' (\epsilon_{\bf p\sigma}-\mu_0)$. Eliminating the impurity distribution function $g$ using Eq.\ (\ref{eq:ImpEq}), we find, in the static limit for the charge current ${\bf J} = {\bf J}_\uparrow + {\bf J}_\downarrow$, the spin current ${\bf J}_{\rm spin} = {\bf J}_\uparrow - {\bf J}_\downarrow$, the heat current ${\bf J}_Q = {\bf J}_{Q \uparrow} + {\bf J}_{Q \downarrow}$, and the spin heat current ${\bf J}_{Q \rm spin} = {\bf J}_{Q \uparrow} - {\bf J}_{Q \downarrow}$, the continuity equations
\begin{eqnarray}
\label{eq:ContinuityEquations}
{\bf \nabla} \cdot {\bf J} &=& 0 \\
{\bf \nabla}  \cdot {\bf J}_{\rm spin}&=& -2 \left(
\tilde \Gamma_\uparrow  \rho_\uparrow
-\tilde \Gamma_\downarrow  \rho_{\downarrow}
+\tilde \Gamma_\uparrow'  \rho_{Q \uparrow}
-\tilde \Gamma_\downarrow'  \rho_{Q \downarrow}\right)
\nonumber \\
{\bf \nabla} \cdot {\bf J}_Q &=&  \epsilon_i \alpha \left(
\Gamma_\uparrow^{\rm sf}  \rho_\uparrow
-\Gamma_{\downarrow}^{\rm sf}  \rho_{\downarrow}
+\Gamma_\uparrow^{\rm sf'}  \rho_{Q \uparrow}
-\Gamma_{\downarrow}^{\rm sf'}  \rho_{Q \downarrow} \right)
\nonumber \\
{\bf \nabla} \cdot {\bf J}_{Q \rm spin}&=&-2 \left(
\hat \Gamma_\uparrow  \rho_{Q\uparrow}
-\hat \Gamma_\downarrow  \rho_{Q\downarrow}
+\hat \Gamma_\uparrow'  \rho_{Q^2 \uparrow}
-\hat \Gamma_\downarrow'  \rho_{Q^2 \downarrow}\right)
\nonumber \\
&&- \epsilon_i \left(
\Gamma_\uparrow^{\rm sf}  \rho_\uparrow
+\Gamma_{\downarrow}^{\rm sf}  \rho_{\downarrow}
+\Gamma_\uparrow^{\rm sf'}  \rho_{Q \uparrow}
+\Gamma_\downarrow^{\rm sf'}  \rho_{Q \downarrow}\right)
\nonumber \\
&&+ 2 \alpha_Q \left(
\Gamma_\uparrow^{\rm sf}  \rho_\uparrow
-\Gamma_{\downarrow}^{\rm sf}  \rho_{\downarrow}
+\Gamma_\uparrow^{\rm sf'}  \rho_{Q \uparrow}
-\Gamma_{\downarrow}^{\rm sf'}  \rho_{Q \downarrow} \right) .
\nonumber
\end{eqnarray}
Here we defined the combined scattering rates $\hat \Gamma_\sigma = \Gamma_\sigma^{\rm sf}+\Gamma_\sigma^{\rm dir}$ and
$\tilde \Gamma_\sigma = \alpha \Gamma_\sigma^{\rm sf}+\Gamma_\sigma^{\rm dir}$ as well as the coefficients $\alpha = {\Gamma_i^{\rm dir}} / ({\Gamma_i^{\rm sf} + \Gamma_i^{\rm dir}})$ and $\alpha_Q = {\Gamma_{iQ}^{\rm sf}}/({\Gamma_i^{\rm sf}+\Gamma_i^{\rm dir}})$. The coefficient $\alpha$ measures the rate of direct relaxation of the impurity spins $\Gamma_i^{\rm dir}=W_{i-}^{\rm dir}-W_{i+}^{\rm dir}$ relative to spin-flip scattering with the conduction electrons,
\begin{equation}
\Gamma_{iQ^n}^{\rm sf} \!\!=\!\! \sum_{\bf pp'}\!
\delta (\!\epsilon_{\bf p\uparrow} \!+\! \epsilon_i \!-\! \epsilon_{\bf p' \downarrow}\!)
(\!\epsilon_{\bf p\uparrow}\!+\!\frac{\epsilon_i}{2}\!-\!\mu_0\!)^n  (f^0_{\epsilon_{\bf p \uparrow}}\!\!\!-\!f^0_{\epsilon_{\bf p' \downarrow}}\!) W^{\rm sf}\!\!,
\end{equation}
i.e., $\alpha$ varies between zero when the impurity spin relaxes only via its interaction with the conduction electrons and unity when direct relaxation dominates. Furthermore, 
\begin{eqnarray}
\rho_{Q^n\sigma} &=& 
\sum_{\bf p} (\epsilon_{\bf p\sigma}-\mu_0)^n \delta n_{\bf p\sigma} 
\nonumber \\
{\bf J}_{Q^n\sigma} &=& 
\sum_{\bf p} (\epsilon_{\bf p\sigma}-\mu_0)^n {\bf v}_{\bf p \sigma} \delta n_{\bf p\sigma} 
\end{eqnarray}
are the electronic densities and currents. Note that the continuity equations contain heat densities or densities of even higher order in energy such as $\rho_{Q^2}$ due to the energy dependence of the spin-flip scattering rates $\Gamma_\sigma^{\rm sf}$ and $\Gamma_\sigma^{\rm dir}$. 

The continuity equations (\ref{eq:ContinuityEquations}) show that the charge current is always conserved while the heat current is conserved only in the absence of inelastic scattering ($\epsilon_i=0$) or of direct relaxation of the impurity spins ($\alpha=0$). Both inelastic spin-flip scattering and direct spin relaxation act as sources or sinks for the spin currents. 

It is instructive to consider first the spin chemical potential in the absence of thermoelectric effects. Then, we can focus on the continuity equations for the charge and spin currents, and neglect the energy derivatives $\Gamma_\sigma'$ of the scattering rates. Inserting Eq.\ (\ref{eq:LinearResponse}) with $S_\sigma=0$, expanding $\rho_\sigma = N_\sigma (\mu_\sigma-\mu_0)$, and rewriting the equations in terms of the charge and spin chemical potentials $\mu_\pm =(1/2)[(\mu_\uparrow - \mu_0) \pm (\mu_\downarrow - \mu_0)]$ yields the coupled diffusion equations
\begin{eqnarray}
\label{eq:PureCharge}
 -\frac{\sigma_+}{e^2} \nabla^2 \mu_+ - \frac{\sigma_-}{e^2} \nabla^2 \mu_- &=& 0\\
 -\frac{\sigma_-}{e^2} \nabla^2 \mu_+ - \frac{\sigma_+}{e^2} \nabla^2 \mu_- &=& -2 \alpha \Gamma_-^{\rm sf} \mu_+ - 2\tilde \Gamma_+ \mu_- \nonumber
\end{eqnarray}
Here, we defined $\sigma_\pm = (\sigma_\uparrow \pm \sigma_\downarrow)$ and $\Gamma_\pm^\alpha = N_\uparrow \Gamma_\uparrow^\alpha \pm N_\downarrow \Gamma_\downarrow^\alpha$. Note that $\Gamma_-^{\rm sf}$ is nonzero only due to the inelastic nature of spin-flip scattering, as becomes evident by the more explicit expression $\Gamma_-^{\rm sf} = [N_\uparrow(\mu_0)N_{\downarrow} (\mu_0 + \epsilon_i) - N_\downarrow(\mu_0) N_{\uparrow} (\mu_0 - \epsilon_i)] ({W^{\rm sf}}/{\sinh \frac{\epsilon_i}{k_B T}})$ \cite{fn1}. For Eq.~(\ref{eq:PureCharge}) one finds a linear and an exponential solution. Determining the spin voltage profile for a ferromagnet of length $L$ with a voltage $\Delta \mu$ applied along the $x$-direction (corresponding to the boundary conditions $\mu_+(x=\pm L/2)=\pm \Delta \mu/2$ and $\mu_-(x=\pm L/2)=0$) one finds for a weak ferromagnet $\epsilon_i,h \ll \epsilon_F$ at small temperatures $k_B T \ll \epsilon_F$ in the bulk of the sample: $\mu_- \simeq - \mu_+ \alpha \Gamma_-^{\rm sf}/\tilde \Gamma_+^{\rm sf}$. This implies, that the linear spatial variation of the charge chemical potential $\mu_+$ due to charge conservation is necessarily accompanied by a linear variation of the spin chemical potential when $\Gamma_-^{\rm sf}$ is nonzero. Interestingly, the linear contribution to $\mu_-$ is entirely determined by the collision integral and thus insensitive to the response coefficients such as the conductivity.

{\it Spin thermoelectrics.}---This mechanism does not carry over directly to spin thermoelectric effects since the latter are generically investigated in the absence of charge currents. Similarly, an ideal spin battery would also be operated without charge currents. Although charge conservation becomes inconsequential under these circumstances, we find that a linear or quasilinear spatial dependence of the spin chemical potential due to inelastic spin-flip scattering persists under quite general circumstances. 

For reference, we first review what happens without inelastic spin-flip scattering ($\Gamma^{\rm sf}=0$), as considered in \cite{Hatami}. Then, both the spin chemical potential and the spin temperature $T_-=(T_\uparrow - T_\downarrow)/2$ exhibit a purely exponential dependence, despite the presence of the conservation law ${\bf \nabla} \cdot {\bf J_Q}=0$ in addition to ${\bf \nabla} \cdot {\bf J}=0$. The reason is that in the equations for the linear modes, $\mu_-$ and $T_-$ decouple completely from $\mu_+$ and $T_+=(1/2)[(T_\uparrow-T_0)+(T_\downarrow-T_0)]/2$. Indeed, expanding the remaining continuity equations (\ref{eq:ContinuityEquations}) around $\mu_0$ and $T_0$ yields
\begin{eqnarray}
\label{eq:ContEqElastic}
{\bf \nabla} \cdot {\bf J}_{\rm spin} &=& -2 ( \Gamma_+^{\rm dir} \mu_- + \gamma  \Gamma_+^{\rm dir'} T_- ) \nonumber \\
{\bf \nabla} \cdot\ {\bf J}_{Q \rm spin} &=& -2 \gamma  \Gamma_+^{\rm dir} T_- 
\end{eqnarray}
where $\gamma = ({\pi^2}/{3}) k_B^2 T_0$, implying that the only sinks and sources of the (heat) spin current are $\mu_-$ and $T_-$. 

As in the absence of thermoelectric effects, spin-flip scattering will mix the equations for the charge and spin components of chemical potential and temperature. This allows for the thermoelectric generation of a linear spin-voltage profile in several circumstances. Expanding the continuity equations (\ref{eq:ContinuityEquations}) around $\mu_0$ and $T_0$ as before, we find, in addition to ${\bf \nabla} \cdot {\bf J}=0$,
\begin{eqnarray}
\label{eq:ExpandedContinuityEquations}
{\bf \nabla} \cdot {\bf J_{\rm spin}} &=& \!
-2 \! \left( \tilde \Gamma_- \mu_+ + \tilde \Gamma_+ \mu_- + \gamma \tilde \Gamma_-' T_+ + \gamma \tilde \Gamma_+' T_- \!\right)\\
{\bf \nabla} \cdot {\bf J_Q} &=& \epsilon_i \alpha \left( \Gamma_-^{\rm sf} \mu_+ + \Gamma_+^{\rm sf} \mu_- + \gamma  \Gamma_-^{\rm sf'} T_+ + \gamma \Gamma_+^{\rm sf'}  T_- \right) \nonumber 
\end{eqnarray}
\begin{eqnarray}
{\bf \nabla} \cdot {\bf J_{Q \rm spin}} &=& 
-2 \gamma \left( T_0 \hat \Gamma_-' \mu_+ + T_0 \hat \Gamma_+' \mu_- + \hat \Gamma_- T_+ + \hat \Gamma_+  T_- \right)  \nonumber \\
&& -\epsilon_i \left(\Gamma_+^{\rm sf}  \mu_+ + \Gamma_-^{\rm sf} \mu_- + \gamma \Gamma_+^{\rm sf'} T_+ + \gamma \Gamma_-^{\rm sf'} T_- \right) \nonumber \\
&&+ 2 \alpha_Q \left(\Gamma_-^{\rm sf} \mu_+ + \Gamma_+^{\rm sf}  \mu_- + \gamma  \Gamma_-^{\rm sf'} T_+ + \gamma \Gamma_+^{\rm sf'}  T_- \right) \nonumber
\end{eqnarray}
Importantly, $\mu_+$ and $T_+$ now also act as sinks and sources for the currents when $\Gamma_-^{\rm sf}$ is nonzero. Consequently, the equations for $\mu_+$ and $T_+$ no longer decouple from those for $\mu_-$ and $T_-$ and the (quasi)linear modes involve also $\mu_-$ and $T_-$. 

A strictly linear spin voltage profile emerges when a second current is conserved in addition to the charge current ${\bf J}$. If direct spin relaxation is negligible for both impurity spins and conduction electrons, the mixing of charge and spin components is accompanied by conservation of the heat and spin currents, ${\bf \nabla} \cdot {\bf J}_Q = 0$ as well as ${\bf \nabla} \cdot {\bf J}_{\rm spin} = 0$ leading to a linear spin voltage in the bulk of the sample (see discussion below). More surprisingly, a linear spin-voltage profile even persists in the presence of direct spin relaxation of the impurity spins. At first sight, this seems unlikely since both the heat current ${\bf J}_Q$ and the spin current ${\bf J}_{\rm spin}$ are no longer conserved. However, a linear combination of the two still satisfies a conservation law, namely
\begin{equation}
{\bf \nabla}\cdot(\epsilon_i {\bf J}_{\rm spin}+2 {\bf J}_Q) = 0 .
\end{equation}
Similarly, a linear spin-voltage profile is possible when there is direct spin relaxation of the conduction electrons, but not of the impurity spins. While the spin relaxation for the conduction electrons spoils the conservation laws for the spin and the spin heat currents, the heat current remains conserved. 

In the full problem where direct spin relaxation is included for both the impurity spins and the conduction electrons, only the charge current is conserved, and strictly speaking, thermoelectric generation of a linear spin-voltage profile becomes impossible. Nevertheless, it is evident from the above discussion that the various currents differ in the robustness of their conservation laws, and consequently some of the length scales for the exponential decays can significantly exceed the spin diffusion length. As a result, one may still find quasilinear spin-voltage profiles in practice even when a linear profile is no longer possible. Specifically, we find from a lengthy but straight-forward analysis that the  three currents ${\bf J}_{\rm spin}$, ${\bf J}_Q$, and ${\bf J}_{Q \rm spin}$ involve the length scales
\begin{eqnarray}
\label{eq:DecayLengths} 
{1}/{\ell_1^2} &\simeq& e^2 \frac{\Gamma_+^{\rm sf}+\Gamma_+^{\rm dir}}{\sigma_+}, \qquad
{1}/{\ell_2^2} \simeq  \frac{1}{\ell_1^2} \frac{\alpha \Gamma_+^{\rm sf}+\Gamma_+^{\rm dir}}{\Gamma_+^{\rm sf}+\Gamma_+^{\rm dir}} \\
{1}/{\ell_3^2} &\simeq& \frac{1}{\ell_1^2} \frac{\alpha}{2} \frac{\epsilon_i \Gamma_+^{\rm dir} \Gamma_-^{\rm sf}}{(\Gamma_+^{\rm sf}+\Gamma_+^{\rm dir})(\alpha \Gamma_+^{\rm sf}+\Gamma_+^{\rm dir})}  \left [ \frac{e}{2\gamma} S_+ - \frac{\Gamma_-^{\rm sf'}}{\Gamma_-^{\rm sf}}  \right] \nonumber 
\end{eqnarray}
with  $S_\pm =(S_\uparrow \pm S_\downarrow)$. Eq.\ (\ref{eq:DecayLengths}) holds for low temperatures, $k_B T \ll \epsilon_F$, and weak ferromagnets, $h, \epsilon_i \ll \epsilon_f$. 

The length $\ell_1$ is the shortest of the three length scales and corresponds to the usual spin diffusion length. The length scale $\ell_2$ is of the same order as the spin diffusion length, except when direct spin relaxation is weak compared to spin-flip scattering on magnetic impurities. In the latter situation, $\ell_2$ becomes much larger than the spin diffusion length and diverges in the absence of direct spin relaxation. Importantly, $\ell_3$ is generically much larger than the spin diffusion length for weak ferromagnets. According to Eq.\ (\ref{eq:DecayLengths}), the ratio $\ell_3/\ell_1$ is of order $(\epsilon_f/\epsilon_i)(\Gamma_+^{\rm dir}/\Gamma_+^{\rm sf})^{1/2} \gg 1$ when direct spin relaxation dominates and $(\epsilon_f/\epsilon_i)(\Gamma_+^{\rm sf}/\Gamma_+^{\rm dir})^{1/2} \gg 1$ in the opposite limit when direct spin relaxation is weak. The length scale $\ell_3$ diverges when no spin relaxation is present for electrons or for impurity spins, as well as in the absence of inelastic spin-flip scattering.

{\it Spin voltage.}---The largeness of $\ell_3$ allows for the thermoelectric generation of a quasilinear spin voltage profile. To illustrate this result and to assess the role of the spin thermopower $S_-$, we consider a ferromagnet of length $L$ with a constant temperature gradient applied along the $x$-direction, similar to the experimental setup in \cite{Uchida}. We then determine the spin-voltage profile from Eqs.\ (\ref{eq:LinearResponse}) and (\ref{eq:ContinuityEquations}) together with the boundary conditions ${\bf J}_{\rm spin} (x = \pm {L}/{2}) = 0$, $T_+ (x = \pm {L}/{2}) = \pm {\Delta T}/{2}$, and $T_- (x = \pm {L}/{2}) = 0$. 

Without direct spin relaxation, $\ell_2$ diverges in addition to $\ell_3$. For low temperatures $k_B T \ll \epsilon_f$ and weak ferromagnets $h,\epsilon_i \ll \epsilon_f$, we then find the spin chemical potential profile
\begin{eqnarray}
\label{eq:NoDirectRelaxation}
\mu_- = -e \left( S_- T_+ + S_+ T_- \right)
\end{eqnarray}
in the bulk of the sample with $T_+ \simeq \Delta T ({x}/{L})$ and $T_- \simeq T_+[(e S_+/2\gamma) \epsilon_i \Gamma_+^{\rm sf}-2{\Gamma_-^{\rm sf}}-\epsilon_i \Gamma_+^{\rm sf'}]/2{\Gamma_+^{\rm sf}}$. Note that both $S_-$ and $S_+$ contribute to the spin chemical potential profile, and it will in general depend on the specifics of the system which of the two terms dominates.

In the general case, with nonzero direct spin relaxation for both the conduction electrons and the impurities, we can focus on the quasilinear contribution decaying on the largest length scale $\ell_3$ and obtain 
\begin{equation}
\mu_-   \simeq  
\alpha \frac{\frac{e}{2} S_+ \Gamma_-^{\rm sf}- \gamma \Gamma_-^{\rm sf'}}{\alpha \Gamma_+^{\rm sf}+\Gamma_+^{\rm dir}} T_+
- \gamma \frac{\alpha \Gamma_+^{\rm sf'}+\Gamma_+^{\rm dir'}}{\alpha \Gamma_+^{\rm sf}+\Gamma_+^{\rm dir}} T_-
\end{equation}
with $T_+ \simeq \Delta T \sinh ({x}/{\ell_3})/2\sinh ({L}/{2 \ell_3})$ and $T_- \simeq T_+[(e S_+/2\gamma) \epsilon_i \Gamma_+^{\rm sf}-2{\Gamma_-^{\rm sf}}-\epsilon_i \Gamma_+^{\rm sf'}]/2{(\Gamma_+^{\rm sf}+\Gamma_+^{\rm dir})}$.
Interestingly, there is no longer a contribution which involves $S_-$. The reason for this is that the contribution of $S_-$ is subdominant in the sense that its spatial profile is controlled by $\ell_2$ rather than $\ell_3$. We therefore find that generically, the spin thermopower does not contribute significantly to the spin Seebeck effect.

{\it Conclusions.}---We have shown that a (quasi)linear spin-voltage profile can be generated thermoelectrically in a ferromagnet with paramagnetic impurities. In the absence of direct spin relaxation of conduction electrons or impurities, the spin chemical potential acquires a strictly linear component due to conservation of a composite current made up from the heat and spin currents. More generally, this conservation law is only approximate and a quasilinear behavior of the spin chemical potential emerges. Surprisingly, the resulting (quasi)linear spin voltage is directly related to the spin thermopower only when neither impurities nor conduction electrons are subject to direct spin relaxation. While it is not clear to which degree our model results apply directly to the recent experiments, they show that a full understanding of the observed quasilinear dependence of the spin chemical potential may also involve purely electronic mechanisms.

The underlying physics of the quasilinear dependence of the spin chemical potential is the spin asymmetry of inelastic spin-flip scattering caused by partially spin-polarized magnetic impurities. Both the spin asymmetry and the inelasticity should also be a feature of scattering of electrons on magnons in itinerant ferromagnets. For this reason, our results suggest that it would be interesting to extend the kinetic theory to include the effects of magnons which however, unlike impurities, have a finite dispersion and propagate.

Our analysis also neglects electron-electron interactions. On the one hand, these will modify the response functions $\sigma_\sigma$, $\kappa_\sigma$, and $S_\sigma$, e.g., by intoducing off-diagonal spin-drag components. In addition, they provide another decay channel for the 
spin heat current which decays on the shortest length scale even in the absence of electron-electron scattering. We therefore expect that electron-electron interactions modify our results only in quantitative detail. We hope to return to these issues in a future publication. 

{\it Acknowledgments.}---We acknowledge financial support through SPP 1285 (Deutsche Forschungsgemeinschaft).


\begin{thebibliography}{99}
\bibitem{fn1} Note that $\epsilon_i/k_BT$ is actually bounded by $1/S$ where $S$ is the impurity spin.
\end{thebibliography}
\end{document}